\documentclass{JINST}

\usepackage{upgreek}
\title{Optimization of the Radiation Hardness of Silicon Pixel Sensors for High X-ray Doses using TCAD Simulations }

\author{ J.~Schwandt$^{a,}$\thanks{Corresponding author.}~,  E.~Fretwurst$^a$, R.~Klanner$^a$,
I.~Pintilie$^{a,b}$ and J.~Zhang$^a$\\
\llap{$^a$}Institute for Experimental Physics, University of Hamburg\\
  Luruper Chaussee 149, D-22761 Hamburg, Germany\\
\llap{$^b$}National Institute of Material Physics,\\
  P.O.Box MG-7, Bucharest-Magurele, Romania\\
  E-mail: \email{joern.schwandt@desy.de}}

\abstract{The European X-ray Free Electron Laser (XFEL) will deliver
27000 fully coherent, high brilliance X-ray pulses per second each with a
duration below 100~fs. This will allow  the recording of diffraction
patterns of single molecules and the study of ultra-fast processes.
One of the detector systems under development for the XFEL is the
Adaptive Gain Integrating Pixel Detector~(AGIPD), which consists of a
pixel array with readout ASICs bump-bonded to a silicon sensor
with pixels of $200 \times 200$~$\upmu $m$^2$.
 The particular requirements for the detector are a high dynamic range
(0, 1 up to $10^5$ 12~keV photons/XFEL-pulse), a fast read-out
and radiation tolerance up to doses of 1~GGy of 12~keV X-rays for 3 years of operation.
At this X-ray energy no bulk damage in silicon is expected. However fixed
oxide charges in the SiO$_{2}$ layer and interface
traps at the Si-SiO$_{2}$ interface will build up.
\\

As function of the 12~keV X-ray dose
the microscopic defects in test structures and the macroscopic
electrical properties of segmented sensors have been investigated.
From the test structures the oxide charge density,
the density of interface traps and their properties as function of dose have been determined.
It is found that both saturate (and even decrease) for doses above a few MGy.
For segmented sensors surface damage introduced by the
X-rays increases the full depletion voltage, the surface leakage
current and the inter-pixel capacitance. In addition an electron
accumulation layer forms at the Si-SiO$_{2}$ interface which
increases with dose and decreases with applied voltage.
Using TCAD simulations with the dose dependent damage parameters obtained
from the test structures the results of the measurements can be reproduced.
This allows the optimization of the sensor design for the XFEL requirements.}


\begin{document}
\section{Introduction}
The European X-ray Free Electron Laser (XFEL), 
currently under construction and planned to be operational in 2015,
will increase the peak brilliance of X-rays by 9 orders of magnitude, 
compared to the third generation storage rings \cite{Altarelli:2007,Graafsma:2009}.
The XFEL will provide 27000 fully coherent X-ray pulses per second each with a duration below 100~fs and with up to
$10^{12}$~12~keV photons. These pulses are distributed in bunch trains containing each 2700 pulses
separated by 220~ns and a bunch train repetition rate of 10~Hz.

The high dynamic range from 0, 1 up to $10^{5}$ 12~keV photons per pixel and the expected radiation 
dose of up to 1~GGy (SiO$_{2}$) for 3~years of operation \cite{Graafsma:2009} 
is a challenge for the design of imaging silicon pixel detectors.

The high density of photons per pulse will create a so called electron-hole plasma
which changes the electric field inside the sensor and will influence among others the linearity,
point spread function and response time of the detector. As shown in \cite{Becker:2010}, for
the reduction of these effects a high operational voltage of the detector is desirable.

The expected radiation dose of up to 1~GGy (SiO$_{2}$) is far beyond today's experience.
The optimization of sensors for this dose demands a good understanding of the radiation damage caused by X-rays.
The aim of this work is to use the relevant parameters for the radiation damage as function of X-ray dose for
the simulation of the sensor performance vs{.} dose, which is needed for the optimization of the sensor design
for radiation hardness.

\section{Formation of X-ray radiation damage and its influence on p${}^{+}$-n sensors}
\label{sec:formandinfl}
For 12~keV photons the main interaction process with Si and SiO$_{2}$ is the photoelectric effect. 
For this photon energy the maximum
energy transfer of the secondary electrons to silicon atoms is 0.011~eV, which is far below the threshold energy of 21~eV
for bulk damages \cite{Akkerman:2001}. Therefore no bulk damage is expected.

In the SiO$_{2}$ the photon-induced damage process is as follow \cite{Barnaby:2006}:
The secondary electrons generate electron-hole pairs (e-h pairs), where the average energy to produce an e-h pair is 18~eV.
Once generated, depending on the electric field, a fraction of the e-h pairs annihilates through recombination.
Due to their high mobility the electrons are rapidly swept out of the dielectric.
The remaining holes will undergo polaron hopping transport via shallow traps in the SiO$_{2}$.
A fraction of these holes may be captured by
deep traps in the oxide bulk or near the Si-SiO$_{2}$ interface,
thereby forming fixed positive charges.
In addition reactions between holes and hydrogen-containing defects or dopant complexes at the interface 
can lead to the formation of interface traps, which have energy levels distributed throughout the silicon band gap.
The details depend among other things on the oxide thickness, electrical field, dose rate, crystal orientation and fabrication technology.

\begin{figure}[htpb]
	\centering
	\includegraphics[width=0.8\textwidth]{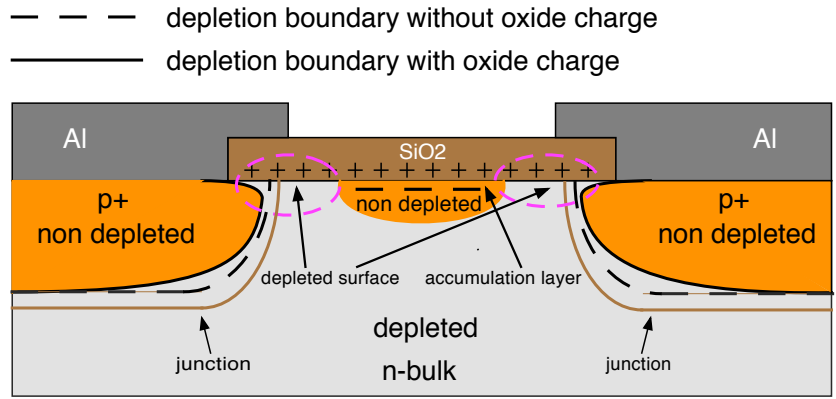}
	\caption{Effects of the oxide charge on a p${}^{+}$-n sensor. A positive oxide charge bends the depletion boundary and increases the width of the
	    accumulation layer. Thus the width of the depleted region at the Si-SiO$_{2}$ interface decreases.} 
	\label{fig:oxideeffects}
\end{figure}

The impact of the surface damage on a segmented  p${}^{+}$-n sensor is shown schematically in figure \ref{fig:oxideeffects}.
Near the junction the positive oxide charge changes the electric field in such a way, that the depletion width along the surface is shorter and 
the depletion boundary exhibits a stronger curvature compared to the case without oxide charges. This results in a local high electric field and
a lower breakdown voltage. In addition, between the junctions below the oxide an electron accumulation layer forms. This
accumulation layer prevents a full depletion of the surface and affects the inter-pixel capacitance, the inter-pixel resistance and
may cause charge losses.  

The effects of the interface traps are twofold. Firstly, depending on the type (donor or acceptor), energy level and Fermi energy
they can be positively or negatively charged and thus contribute to the effective oxide charge.
Secondly, due to surface recombination at the depleted surface they are responsible for the surface leakage current, which can be
parametrized as $I_{s} =  0.5 q_{0} s_{0} n_{i} A_{s}$ with the elementary charge $q_{0}$, 
the intrinsic carrier density $n_{i}$, the depleted surface $A_{s}$ and
the surface recombination velocity $s_{0} = \sigma v_{th} N_{it}$, where $\sigma$ is the capture cross section, 
$v_{th}$ the thermal velocity of the carriers and $N_{it}$ the interface trap density. 

\section{Measurement results for X-ray induced radiation damage}
\begin{figure}[htpb]
	\centering
	\includegraphics[width=0.8\textwidth]{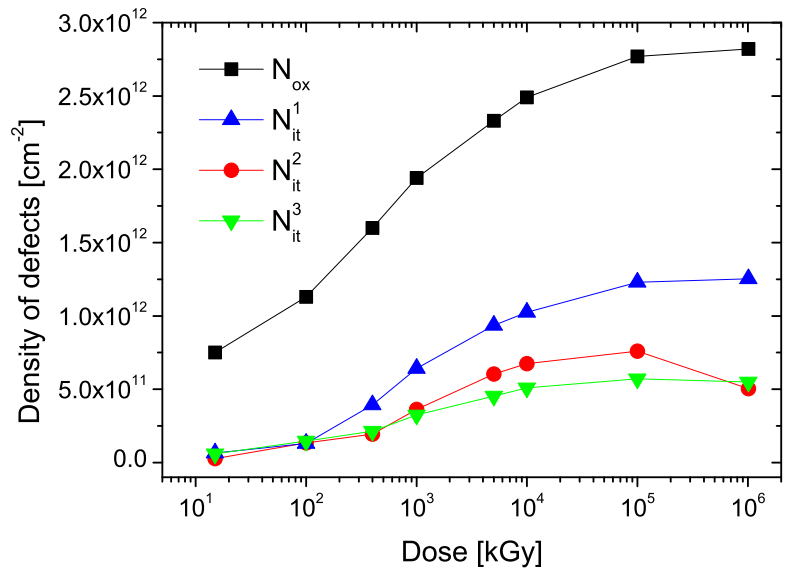}
	\caption{Dose dependence of the fixed oxide charge density $N_{ox}$ and the densities of the
           three dominant X-ray induced interface traps as function
	  of X-ray irradiation dose after annealing at 80 $^\circ$C for 10 minutes and crystal orientation <100>. }
	\label{fig:NoxNitdosedep}
\end{figure}

From the previous discussion it follows that the relevant parameters are
the oxide charge density $N_{ox}$ and the interface trap density distribution.
For their experimental determination MOS capacitors and gate controlled diodes fabricated by CiS were irradiated
up to 1~GGy with 12~keV X-rays and Thermally Dielectric Relaxation Current (TDRC), 
Capacitance/Conductance-Voltage (C/G-V) measurements as function of frequency have been performed. 
Details on the irradiation setup, the test structures and the extraction of the damage parameters
can be found in \cite{Zhang:2011a} and \cite{Zhang:2011b}.
Here, only the summary is given: The radiation induced effects can be described by a dose dependent value of the fixed oxide charge
density $N_{ox}$ and the densities of three acceptor like interface trap levels $N_{it}^{1,2,3}$.  
For MOS capacitors with crystal orientation <100> the dose dependences 
of $N_{ox}$ and of $N_{it}^{1,2,3}$ are shown in figure \ref{fig:NoxNitdosedep}. 
As function of dose they
increase and finally saturate for values above 10 to 100 MGy. The saturation value of $N_{ox}$ is $2.8\times10^{12}$~cm$^{-2}$.
Similar results were obtained for gate controlled diodes with <111> orientation.
In the TCAD simulations the damage parameters presented in table \ref{tab:useddose} were used.
The interface trap densities $N_{it}^{1,2,3}$ were not explicitly introduced in the simulation, instead the measured
surface recombination velocity $s_{0}$ was used. 
Its value as function of dose was obtained from I/V measurements of gate controlled diodes \cite{Perrey:2011}.
\begin{table}[htpb]
	\begin{center}
		\begin{tabular}{|c||c|c|c|c|c|}
		        \hline
			Dose [MGy] & 0 & 0.1& 1 & 10 & 100 \\ \hline
			$N_{ox}$ [cm$^{-2}$] & $1.0\times10^{11}$  & $1.3\times10^{12}$  & $2.1\times10^{12}$ & $2.8\times10^{12}$ &
			$2.9\times10^{12}$  \\ \hline
			$s_{0}$ [cm/s] & 8 & $3.5\times10^{3}$  & $7.5\times10^{3}$ & $1.2\times10^{4}$ & $1.1\times10^{4}$  \\
			 \hline
		\end{tabular}
		\caption{Dose dependence of the fixed oxide charge density $N_{ox}$ and the surface recombination velocity $s_{0}$ 
		  measured on gate controlled diodes with <111> orientation after annealing at 80 $^\circ$C for 60 minutes.}
		\label{tab:useddose}
	\end{center}
\end{table} 

\section{TCAD simulations for the AGIPD sensor} 
The Adaptive Gain Integrating Pixel Detector (AGIPD) \cite{Henrich:2011} is one of the three 
large 2D detector projects approved by the European XFEL Company.
Based on detector and science simulations \cite{Potdevin:2009}, studies on plasma effects and technology criteria,
the AGIPD collaboration has specified the sensor parameter listed in the table \ref{tab:sensorspec}.
\begin{table}[htpb]
	\begin{center}
	\begin{tabular}{|c||c|}
		        \hline
          		Parameter & Specification \\ \hline \hline
			Thickness & 500 $\upmu$m \\ \hline
			Pixel size & 200 $\upmu$m $\times$ 200 $\upmu$m \\ \hline
			Type & p$^+$-n \\ \hline
			Resistivity & ~5 k$\Omega\cdot$cm \\ \hline
			$V_{fd}$ & < 200 V \\ \hline
			$V_{op}$ & $\geqslant$ 500 V \\ \hline
			$C_{int}$ & < 0.5 pF \\ \hline
			$I_{leak}$ & 1 nA/pixel \\ 
			 \hline
        \end{tabular}
        \caption{AGIPD sensor specifications ($V_{fd}$ full depletion voltage, 
                     $V_{op}$ operation voltage, $C_{int}$ inter-pixel capacitance, $I_{leak}$ leakage current).}
        \label{tab:sensorspec}
	\end{center}
\end{table}       

\begin{figure}[htpb]
	\centering
	 \includegraphics[width= 0.8\textwidth]{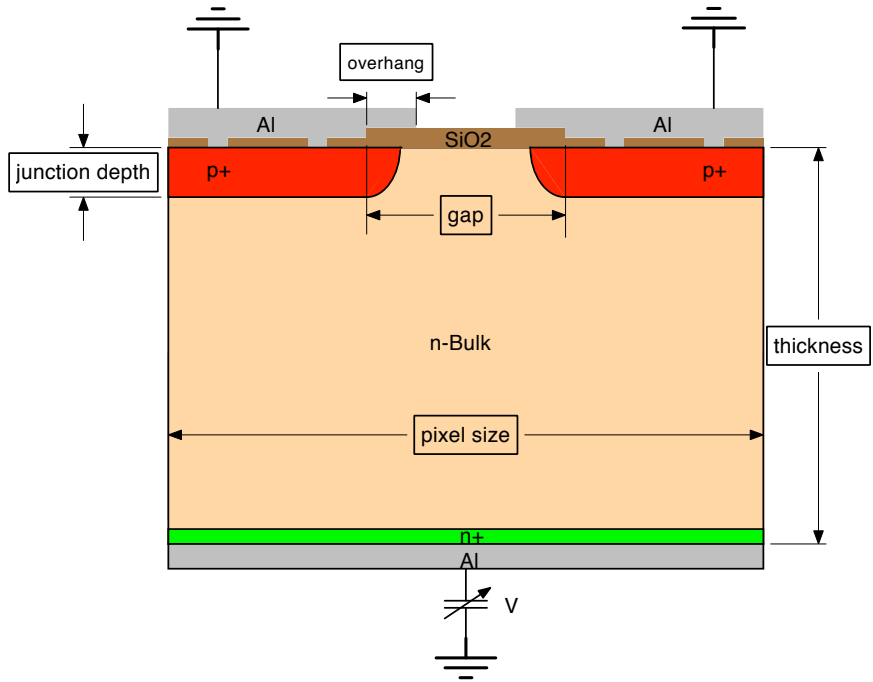}
           \caption{Sketch of sensor region simulated. The parameters optimized as function
            of the voltage applied between pixel (p$^{+}$) and backplane (n$^{+}$)
            in view of radiation hardness are: Al-overhang, gap-size and junction depth. }
           \label{fig:sensorsketch}
\end{figure}

The parameters which can been optimized in the design of the sensor to fulfill the specifications (see figure \ref{fig:sensorsketch})
are the inter-pixel gap, the metal overhang, 
the curvature at the implant corners and the guard ring structure.
Other parameter like oxide thickness, junction depth and passivation thickness are typically
given by the vendors' technology and may only be changed in a limited range or not at all.

To determine the optimum values for above parameters taking into account the surface damage
Synopsys TCAD simulations \cite{Synopsys} were performed. In a first step the focus was put on the optimization of the gap and the
metal overhang. For this 2D simulations were carried out and the resulting currents and capacitances were scaled to 3D.
Using 2D simulations, even if the problem of pixels is intrinsically a 3D one, is motivated by the fact, that the inclusion of
surface damages requires a very fine mesh at the Si-SiO$_{2}$ interface,
which requires large amount of computer time and direct access memory.
 
To take the dose dependent effects into account a similar oxide thickness as the one of the test structures was used
and the measured values for the fixed oxide charge densities together with the surface recombination
velocities given in table \ref{tab:useddose} were used.
For the fixed oxide charge the assumption was made, that the spatial distribution
along the Si-SiO$_{2}$ interface is uniform.
The effect of the interface traps is taken into account only via the surface recombination velocity. 
 
\subsection{Doping profiles}
As discussed in section \ref{sec:formandinfl} 
the doping profile near the implant edges is essential for the simulation of the effects of surface damages.
To get realistic doping profiles 2D process simulations were performed involving the oxide deposition,
etching of the implant window, ion implantation and the dopant activation steps.
The details are: Phosphorous doping of $10^{12}$~cm$^{-3}$ and <111> orientation for the wafer
and boron doses of $10^{15}$~cm$^{-2}$ to $10^{16}$~cm$^{-2}$ and energies
from 70~keV to 200~keV for the p$^{+}$ implantation. 

\begin{figure}[htpb]
	  \includegraphics[width=0.5\textwidth]{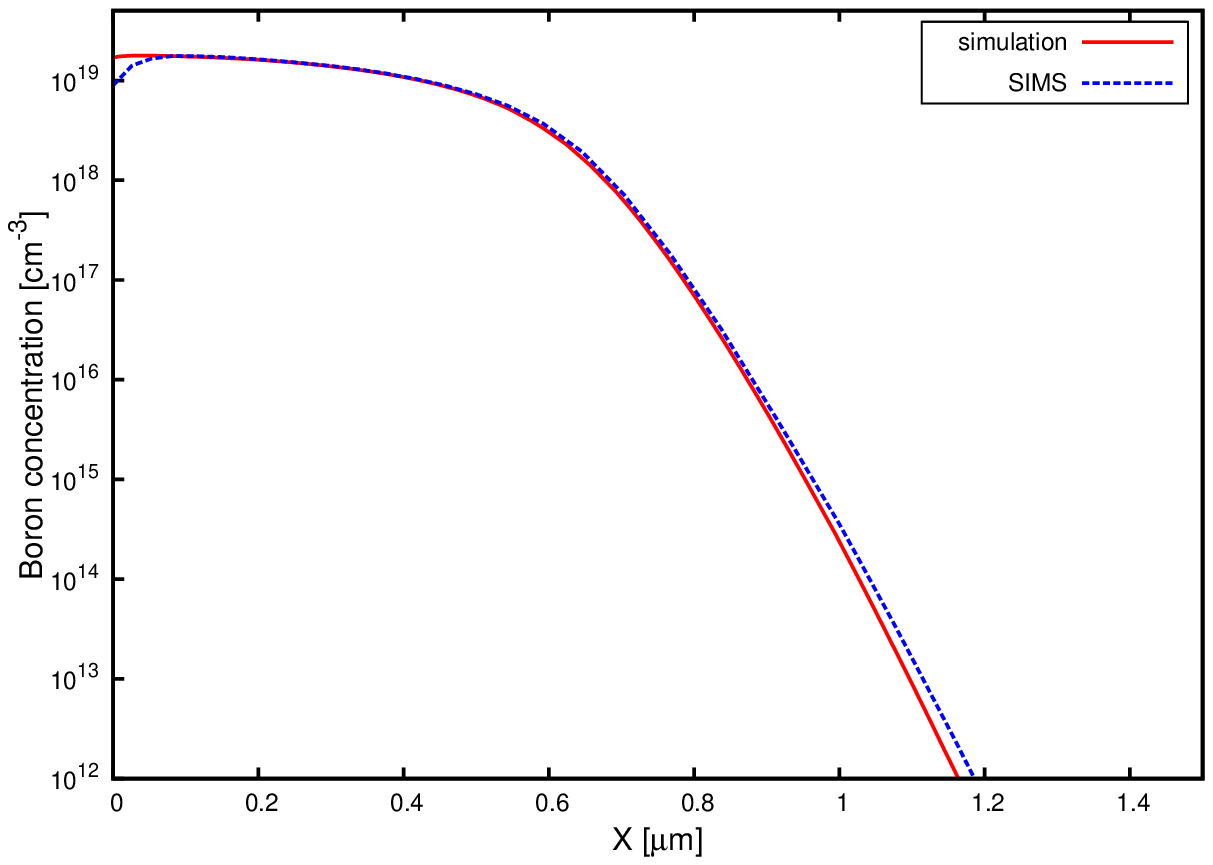}\hfill
	  \includegraphics[width=0.55\textwidth]{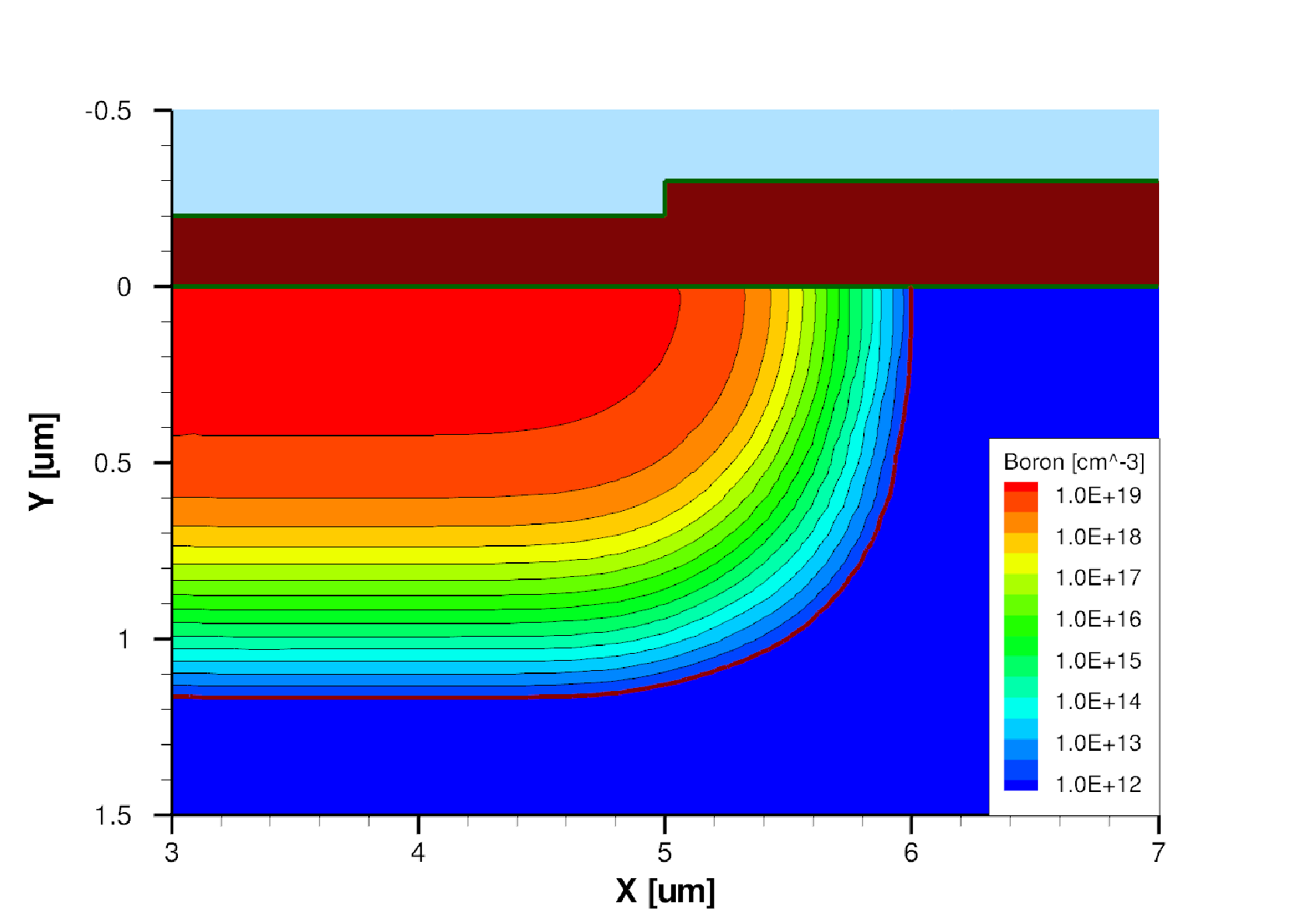}	   
           \caption{Left: Comparison of the simulated boron profile with a parameterization of the SIMS measurement.
                        Right: Simulated 2D boron profile.}
           \label{fig:boronimpl}
\end{figure}

The simulation was calibrated by comparison of the boron depth profile 
for the dose of $10^{15}$~cm$^{-2}$ and the energy of 70~keV 
with a parameterization of the SIMS measurement for the same process, 
as shown in the left plot of figure \ref{fig:boronimpl}.
The right plot of figure \ref{fig:boronimpl} shows the corresponding 2D boron profile from which one
finds a junction depth of 1.2~$\upmu$m 
and a lateral extension of 1.0~$\upmu$m, measured from the edge of the implant.
In the following device simulations this doping profile was used and for comparison, the profile from
a process which gives a junction depth of 2.4~$\upmu$m and a lateral extension of 1.95~$\upmu$m.

\subsection{Models used in device simulation and scaling from 2D to 3D}
The geometric model used in the 2D device simulation is shown in figure \ref{fig:sensorsketch}.
The simulated pixel size was 200~$\upmu$m,  the sensor thickness 500~$\upmu$m and the oxide thickness 300~nm.
The gap and metal overhang were varied in the range given in table \ref{tab:geometries}.
\begin{table}[htpb]
	\begin{center}
		\begin{tabular}{|c||c|c|c|}
		        \hline
			gap [$\upmu$m] & 20 & 30 & 40  \\ \hline
			overhang  [$\upmu$m] & 0, 2.5, 5 & 5, 10 & 0, 2.5, 5, 10 \\
			 \hline
		\end{tabular}
		\caption{Simulated gap and metal overhang values}
		\label{tab:geometries}
	\end{center}
\end{table}

The device simulations were performed for a temperature of 293~K using the drift-diffusion model
with standard models, as implemented in \cite{Synopsys}, for mobility (with degradation at interfaces),
SRH recombination (carrier lifetime 1~ms), band-to-band tunneling 
and avalanche generation, where the van Overstraeten - de Man model was used.  
At the top of the SiO$_{2}$ Neumann boundary conditions were used.

To scale the 2D results to 3D the following simplifying assumptions were made:
For the bulk current, which dominates the current of a non-irradiated sensor, a scaling with the bulk volume was performed.
For the surface current, which dominates for irradiated sensors, a scaling with the SiO$_{2}$ area was performed. This results in an
additional factor 2 compared to the bulk scaling.
For the scaling of the 2D simulated inter-strip capacitances $C^{str}_{int,sim}$
analytical expressions for the inter-strip capacitance $C^{str}_{int,theo}$ \cite{Cattaneo:2010} and 
for the inter-pixel capacitance $C^{pix}_{int,theo}$ \cite{Cerdeira:1997},
which both neglect the metal overhang, were used and the assumption 
that the 3D simulated inter-pixel capacitance $C^{pix}_{int,sim}$ is given by
\begin{equation}
C^{pix}_{int,sim}= \frac{C^{str}_{int,sim}}{C^{str}_{int,theo}} \times C^{pix}_{int,theo}
\end{equation}
was made.

\subsection{Simulation results}
\begin{figure}[htpb]
	  \includegraphics[width=0.5\textwidth]{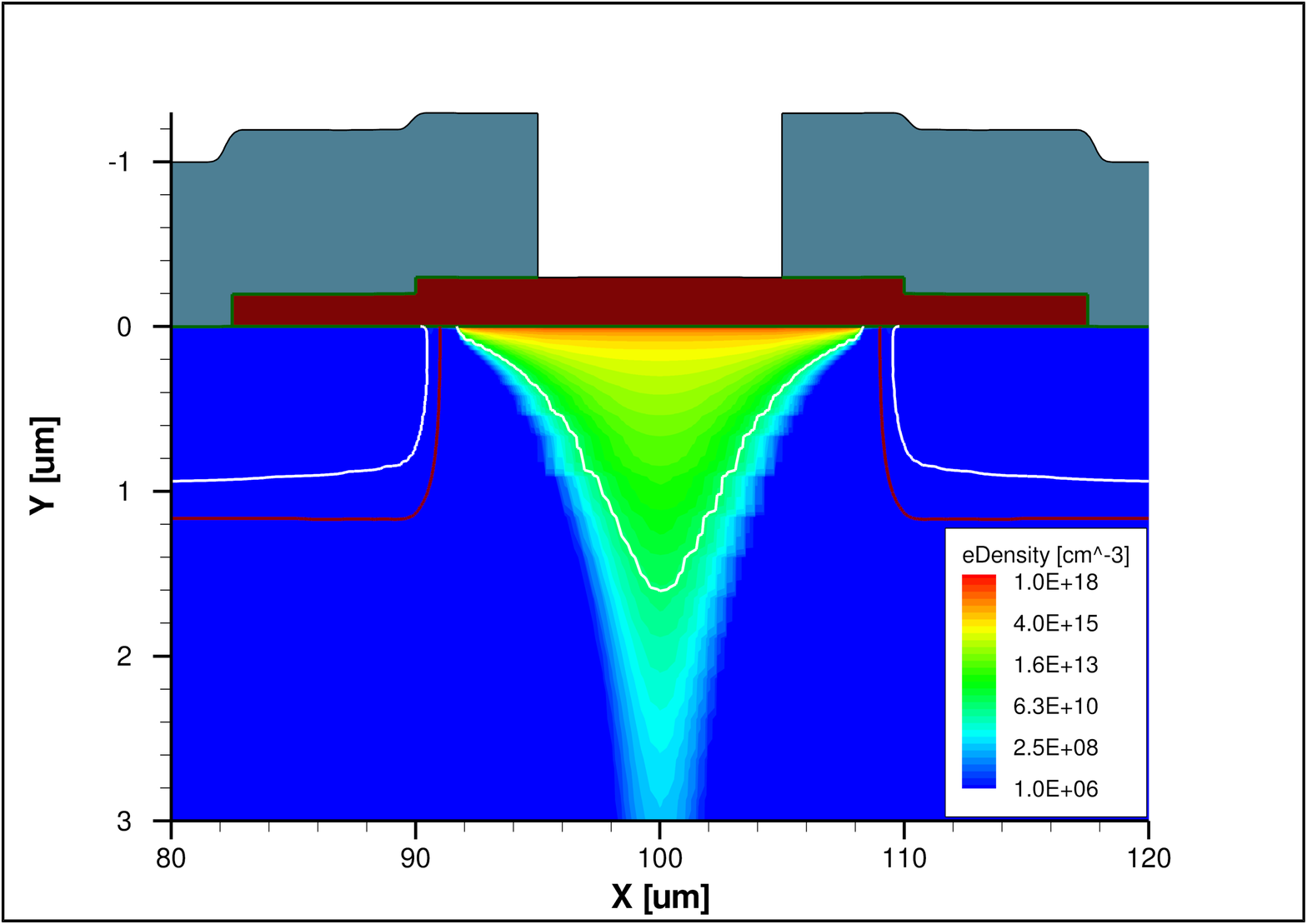}\hfill	   
	  \includegraphics[width=0.5\textwidth]{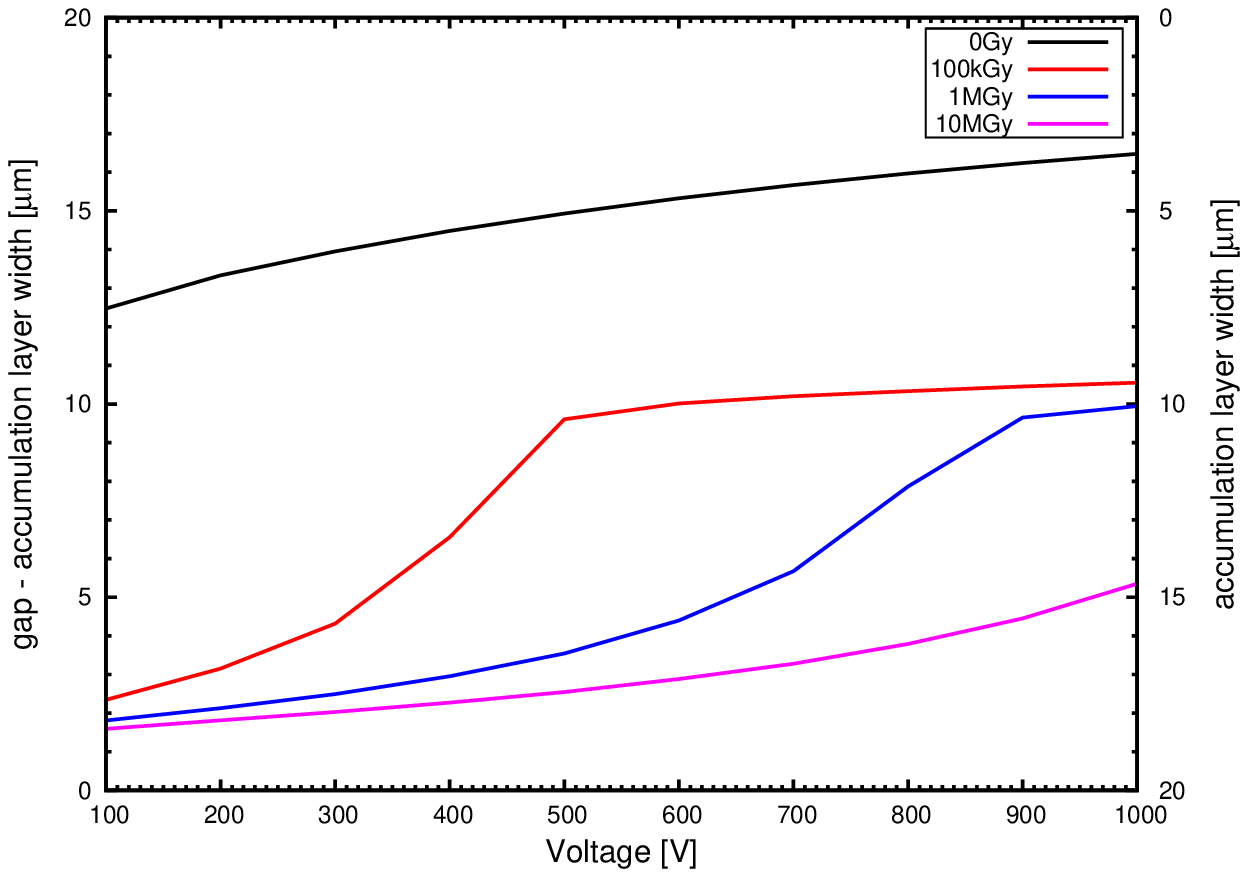} 
          \caption{Left: Electron density for a sensor with 20~$\upmu$m gap and 5~$\upmu$m metal overhang at 500V and 1MGy.
                        Right: Difference of the width of the gap minus the width of the accumulation layer vs{.} voltage. The scale on the
                        right side gives the accumulation layer width.}
           \label{fig:accuml}
\end{figure}

First the results of the dose and voltage dependence of the electron accumulation layer are presented. 
They were obtained from the electron density 20~nm below the Si-SiO$_{2}$ interface using
as condition a value of the electron density equal to the bulk doping.
Figure \ref{fig:accuml} left shows the results of a 2D-simulation for a sensor
with 20~$\upmu$m gap, 5~$\upmu$m metal overhang and radiation effects corresponding 
to 1~MGy for a 500~$\upmu$m thick sensor biased to 500~V.

The voltage dependence for different dose values is shown in the right plot of figure \ref{fig:accuml} 
for sensor with a 20~$\upmu$m gap, 5~$\upmu$m metal overhang and 1.2~$\upmu$m junction depth.
The difference between the gap width and the accumulation layer width is plotted, 
since the surface current is proportional to this difference.
As can be seen, in the non-irradiated case a small (3.5 - 7.5~$\upmu$m), only weakly voltage dependent
accumulation layer is present. In the case of 100~kGy and 1~MGy at low voltages practically the entire
region between the junctions is covered by the accumulation layer and for high voltages the region under the metal
overhang depletes. For 10~MGy  the breakdown voltage is 494~V, 
where as criterion for breakdown an
electron or hole ionization integrals equals to one was used \cite{Sze}.

\begin{figure}[h]
	  \includegraphics[width=0.5\textwidth]{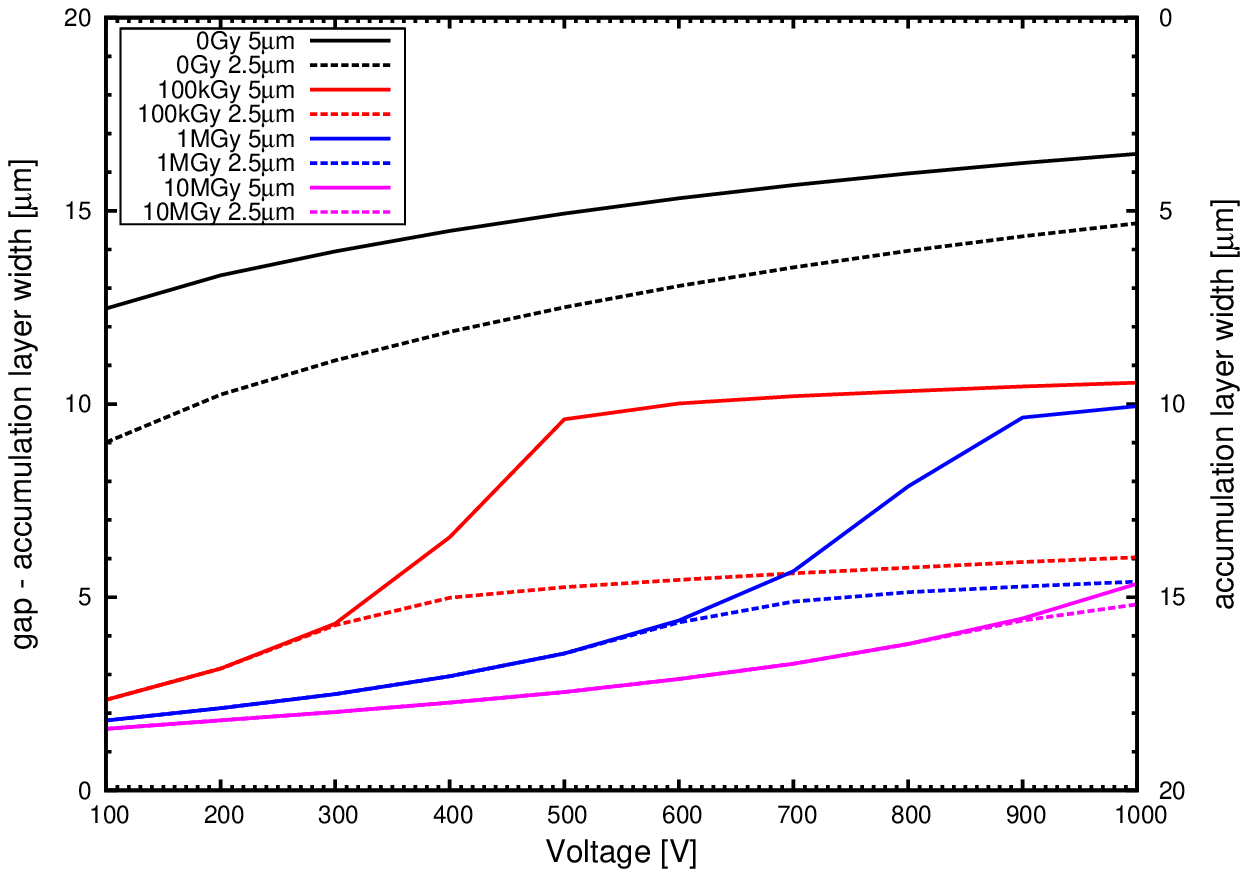}\hfill	   
	  \includegraphics[width=0.5\textwidth]{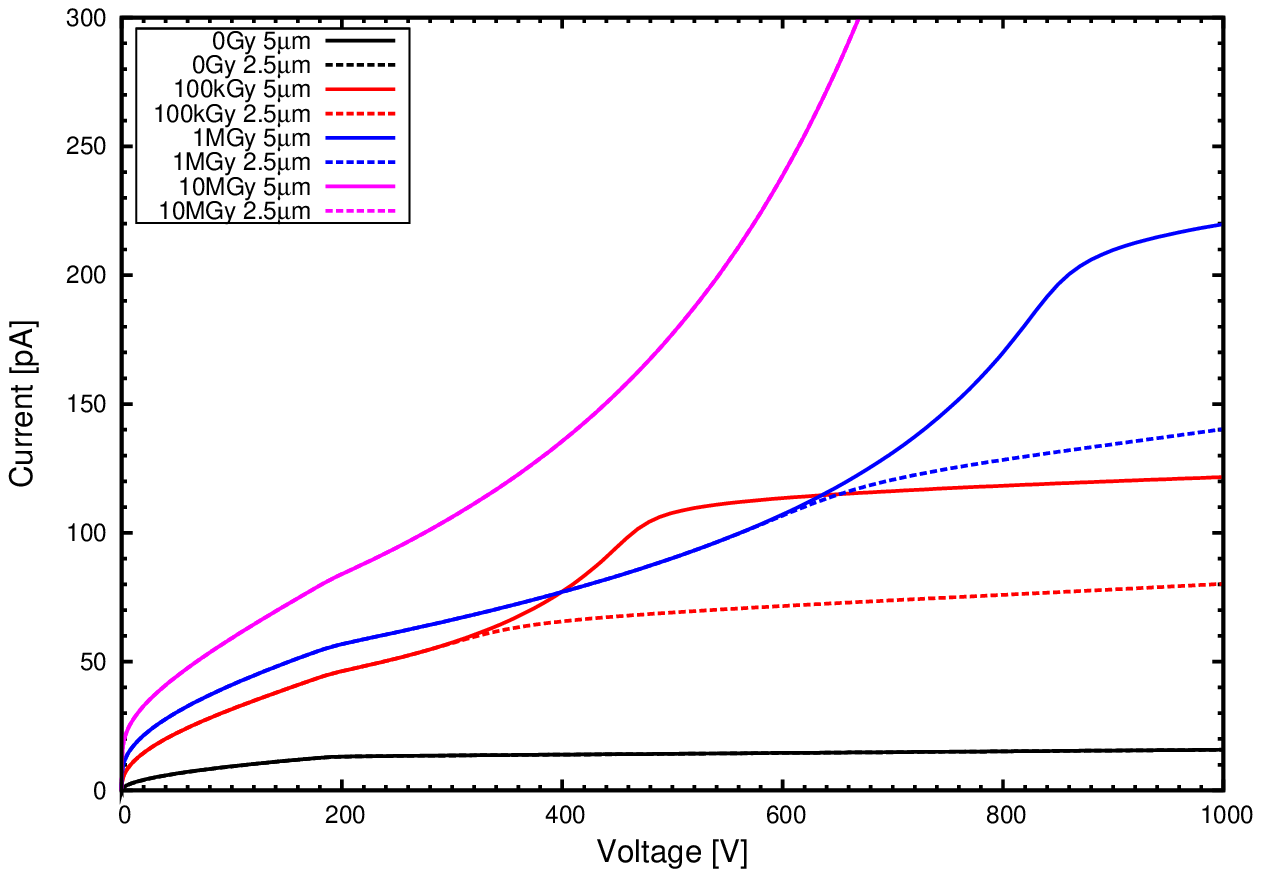} 
          \caption{Left: Difference of the width of the gap minus the width of the accumulation layer vs{.} voltage 
                        for a sensor with 2.5 and 5~$\upmu$m metal overhang.
                       Right: Current vs{.} voltage for 2.5 and 5~$\upmu$m metal overhang.}
           \label{fig:overhang}
\end{figure}

The effect of different overhang values, 2.5 and 5~$\upmu$m, for the same gap and junction depth is shown 
 in figure \ref{fig:overhang}, where in addition to the difference gap-accumulation layer width the I-V curves are plotted. 
 For irradiated sensors at sufficiently high voltages the accumulation layer for the 5~$\upmu$m overhang 
 is smaller than for the 2.5~$\upmu$m resulting in a higher current for the larger overhang. 
 In the 10~MGy case the breakdown voltage of 494~V is the same for both overhang values.
 Without metal overhang the breakdown voltage is much lower (296~V).
 The 20~$\upmu$m gap is the smallest which was simulated. Therefore for all
 larger gap sizes the breakdown voltage will be lower. 

\begin{figure}[htpb]
	\centering
	  \includegraphics[width=0.5\textwidth]{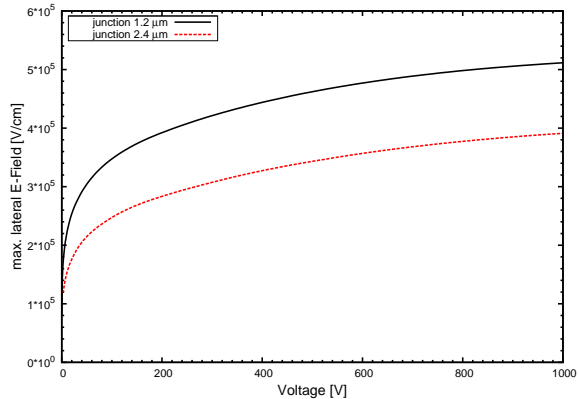}\hfill	   
          \caption{Maximum lateral electric field in silicon as function of voltage for sensors with a junction depth of 1.2~$\upmu$m and 2.4~$\upmu$m
                        (20~$\upmu$m gap, 5~$\upmu$m overhang, 10~MGy).}
           \label{fig:maxlate}
\end{figure}

The maximum electric field in the silicon for a given oxide thickness can only be reduced with
a deeper junction, since this leads to a smaller curvature at the implant edges.
For the 10~MGy and 20~$\upmu$m gap,
figure \ref{fig:maxlate} shows the comparison of the maximum lateral electric field in the silicon
as function of voltage for the 1.2~$\upmu$m and 2.4~$\upmu$m deep junctions.
A reduction of the electric field is seen:
At 500~V  the electric field of the 2.4~$\upmu$m deep junction is about 30\% lower compared
to the 1.2~$\upmu$m junction. 
The breakdown voltage for the deep junction is in this case above 1000~V. Breakdown voltages above
1000~V were also obtained for
a gap of 30~$\upmu$m with 5 and 10~$\upmu$m overhang 
and 40~$\upmu$m gap with 10~$\upmu$m overhang.
From the  I-V curves it is found, that the current of all investigated cases are within the specifications.

\begin{figure}[htpb]
	  \includegraphics[width=0.5\textwidth]{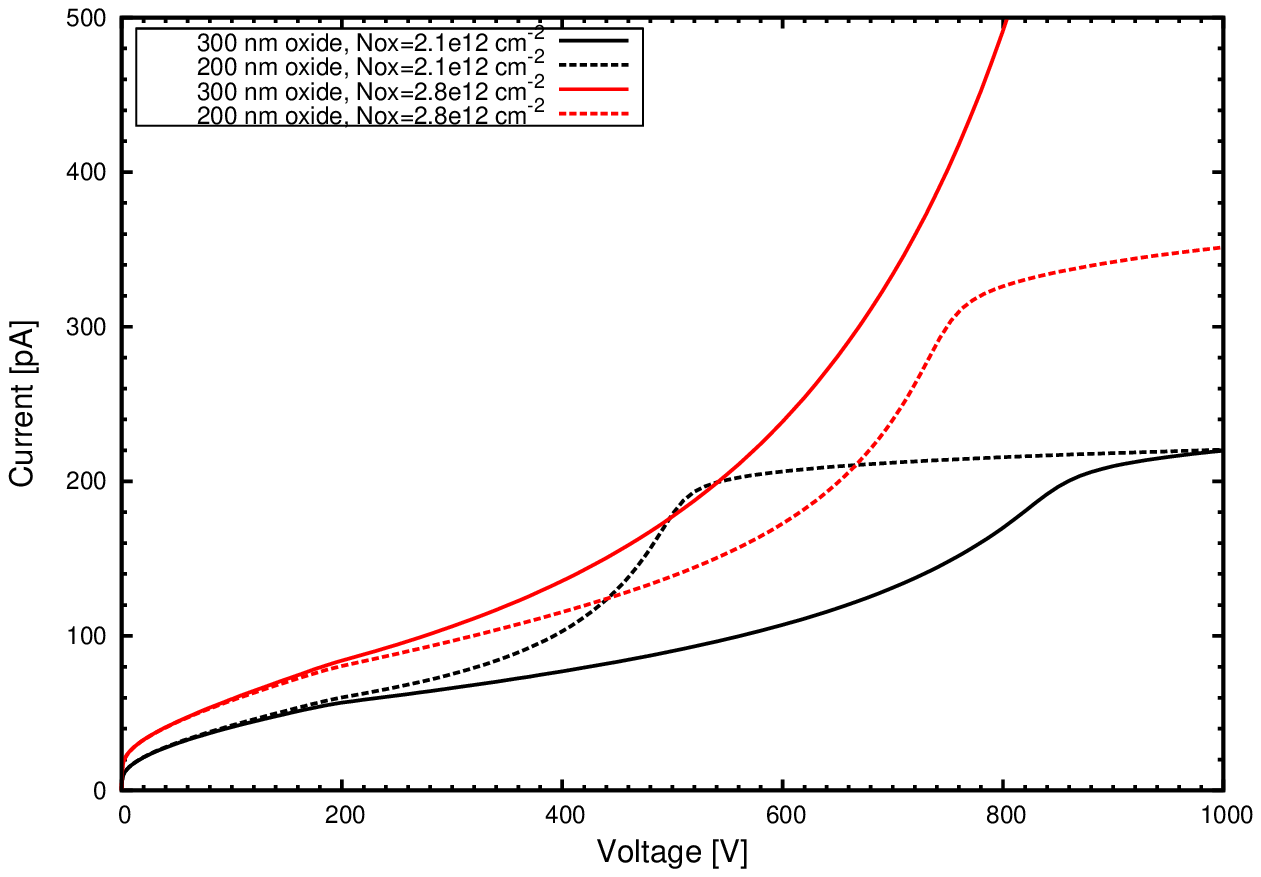}\hfill	   
	  \includegraphics[width=0.5\textwidth]{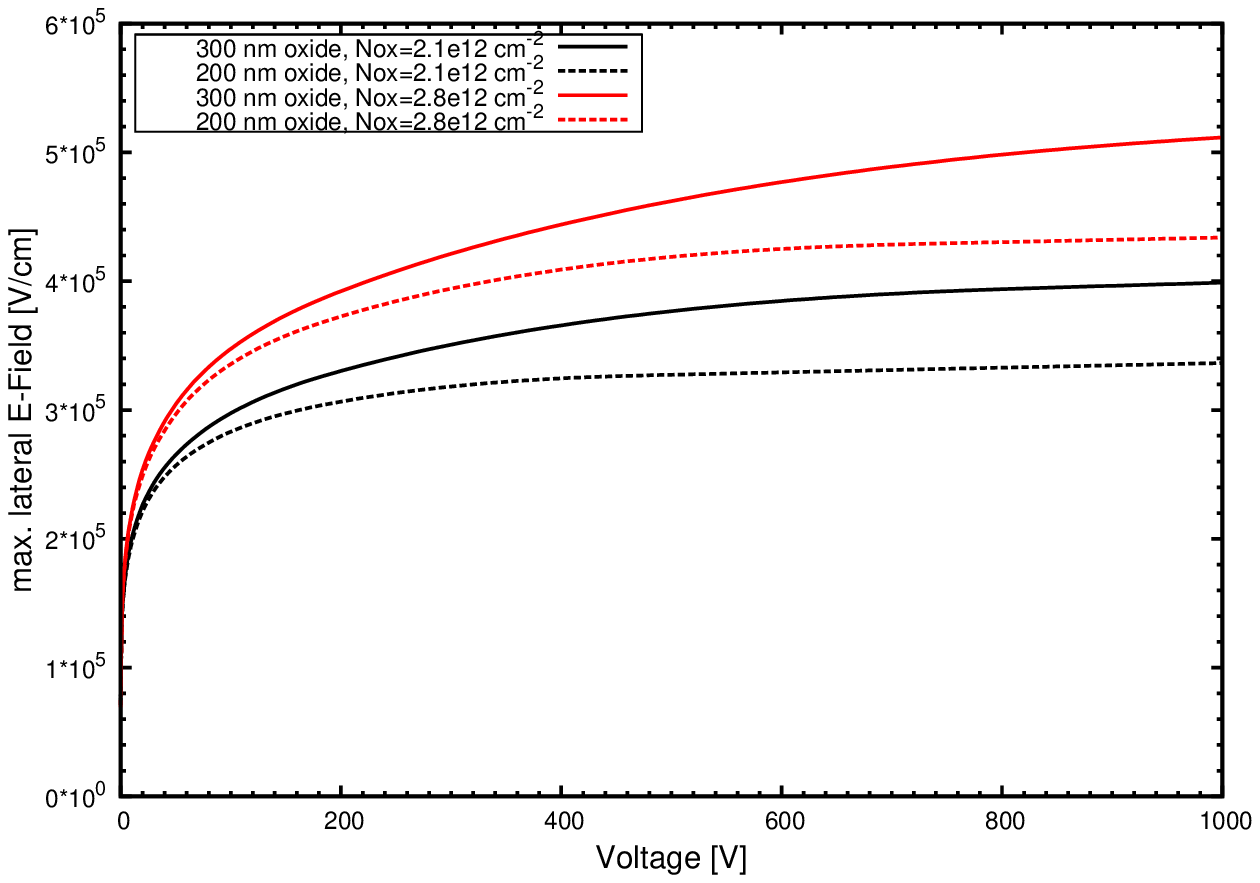} 
          \caption{Left: Current vs{.} voltage for a sensor with 200~nm and 300~nm thick oxide (20~$\upmu$m gap, 5~$\upmu$m overhang).
                       Right: Maximum lateral electric field in silicon as function of voltage for a sensor with 200~nm and 300~nm thick oxide.
                           (20~$\upmu$m gap, 5~$\upmu$m overhang).}
           \label{fig:oxideth}
\end{figure}

Another possibility to reduce the maximum electric field inside the silicon is
the reduction of the oxide thickness as can be seen in the right 
plot of figure \ref{fig:oxideth}, where the maximum lateral electric field in the silicon
as function of voltage for an oxide thickness of 200 and 300~nm is plotted.
The simulations are again performed for the 20~$\upmu$m gap, 5~$\upmu$m overhang and the 1.2~$\upmu$m deep junction.
In addition the assumption was made, that the amount of fixed oxide charges are the same in both cases.
A fixed oxide charge density of $N_{ox} = 2.1\times10^{12}$~cm$^{-2}$ corresponds to a dose of 1~MGy
and $N_{ox} = 2.8\times10^{12}$~cm$^{-2}$ corresponds to a dose of 10~MGy for a 300~nm thick oxide.

The reduction of the field at 500~V  is about 15\% for the lower and about 8\% for the higher oxide charge.
From the I-V curves, left plot of figure \ref{fig:oxideth}, it can be seen, that for the thinner oxide
the region under the metal overhang depletes at lower voltages and the breakdown is above 1000~V.

To compare the results of the 2D (strip) simulations with 3D (pixel) simulations
a quarter of a pixel with a gap of 20~$\upmu$m and 5~$\upmu$m overhang was simulated. 
Due to difficulties in including the doping profile from the process simulation a 
Gaussian profile with a junction depth of 1.5~$\upmu$m was used.
The number of mesh points was $4.4\times10^{5}$ and $9.5\times10^{5}$ for the 3D case, where
approximately 80\% of the grid points were used for the refinement of the mesh at the Si-SiO$_{2}$ interface.
The total cpu time for the simulations using the smaller mesh were, depending on the dose, 180~h to 406~h.

\begin{figure}[htpb]
	\centering
	  \includegraphics[width=0.5\textwidth]{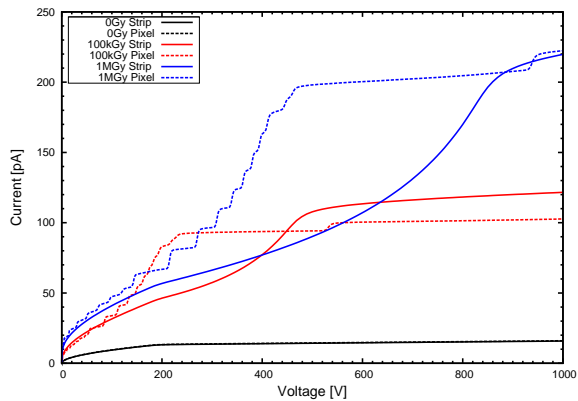} 
          \caption{Comparison of the I-V characteristics between 2D (strip) and 3D (pixel) 
          simulation for different dose values (20~$\upmu$m gap, 5~$\upmu$m overhang).}
           \label{fig:2dvs3d}
\end{figure}

In figure \ref{fig:2dvs3d} the I-V curves for the smaller mesh for different dose values together with 
the results of the 2D simulations are shown. The steps in the I-V curves in the 3D simulations are related
to the mesh of the Si-SiO$_{2}$ interface below the metal overhang. 
Using the finer mesh gives similar I-V curves, only the steps were reduced.

From the comparison of the I-V curves of the 2D and 3D simulations one finds, that
for low and high voltages the simple scaling of the geometry gives reasonable results. 
In the 3D simulation however, the depletion
under the metal overhang appears at a lower voltage, 
which causes in the difference of the voltage dependence of the 2D and 3D simulated current.

To verify that the specification of the inter-pixel capacitance can be met with the simulated geometries,
2D capacitance simulations were performed. For all geometries and doses the full depletion voltage is in the
range of 188-194~V and the capacitance of a pixel to the backplane is 8.5~fF.
In table \ref{tab:capac} the simulated inter-pixel capacitances for a junction depth of 1.2~$\upmu$m 
and a overhang of 5~$\upmu$m are given for a dose of 0~Gy and 1~MGy.
The values for the capacitance are given at full depletion (190~V) and at 500~V.
Since, the accumulation layer decreases with voltage also the inter-pixel capacitance decreases.
The reason why at the depletion voltage the interstrip capacitance at 1~MGy for the 40~$\upmu$m gap
is larger then for the 30~$\upmu$m gap, is still under study.
All inter-pixel capacitances are well within the specifications (< 0.5~pF). 
\begin{table}[htpb]
	\begin{center}
		\begin{tabular}{|c|c|c|c|c|c|c|c|}
		     \hline
		          \multicolumn{2}{|c|}{ } &  \multicolumn{3}{|c|}{ V = 190 V } &  \multicolumn{3}{|c|}{ V = 500 V }\\
		        \hline
		        Gap & Dose & acc. layer & 2D C$_{int}$ & 3D C$_{int}$ & acc. layer &  2D C$_{int}$  & 3D C$_{int}$  \\
		          $\upmu$m & MGy & [$\upmu$m] & [fF/$\upmu$m] & [fF] & [$\upmu$m] & [fF/$\upmu$m] & [fF]  \\ \hline \hline
		            \raisebox{-1.5ex}[1.5ex]{20} & 0 & 6.7  & 0.12 & 96 & 5.1 & 0.12 & 93 \\ \cline{2-8} 
		             & 1 & 17.9  & 0.39 & 305 & 16.5  & 0.33 & 259 \\  \hline
		    \raisebox{-1.5ex}[1.5ex]{30} & 0 & 13.3 & 0.10  & 73 & 10.3 & 0.10 & 71 \\\cline{2-8} 
		             & 1 & 26.8 & 0.17 & 117 & 20.1 & 0.13 & 92 \\ \hline
		      \raisebox{-1.5ex}[1.5ex]{40} & 0 & 19.8 & 0.09 & 59 & 15.4 & 0.09  & 56 \\ \cline{2-8} 
		             & 1 &  34.3 & 0.25 & 163 & 29.5  & 0.13 & 82 \\ 
			 \hline
		\end{tabular}
		\caption{Calculated inter-pixel capacitance for 1.2~$\upmu$m deep junction and 5~$\upmu$m overhang. }
		\label{tab:capac}
	\end{center}
\end{table} 

\section{Summary}

Using MOS capacitors and gate controlled diodes
the oxide charge density, the density of interface traps and their properties
were determined as function of dose up to 1~GGy.
These dose dependent parameters were used in the TCAD simulation
to predict and optimize the performance of sensors designed for the AGIPD project.

From the simulation it is found, that the specification of dark current, 
inter-pixel capacitance and full depletion voltage can be met.

\acknowledgments{This work was done within the Project \emph{Radiation Damage} financed by the European XFEL-Company in close collaboration with the AGIPD project. Additional support was provided by the Helmholtz Alliance \emph{Physics at the Terascale}.
}

\end{document}